\documentclass[12pt,a4paper]{article}
\usepackage{amsmath, amssymb, fullpage, graphics}
\usepackage{caption, graphicx, fancyhdr, fancybox}
\usepackage[cp866]{inputenc}
\usepackage{color}

\newcommand{\bv}{\mathbf{v}}

\newcommand{\bU}{\mathbf{U}}
\newcommand{\bF}{\mathbf{F}}
\newcommand{\bA}{\mathbf{A}}
\newcommand{\bB}{\mathbf{B}}
\newcommand{\bC}{\mathbf{C}}
\newcommand{\bI}{\mathbf{I}}
\newcommand{\bP}{\mathbf{P}}
\newcommand{\bxi}{\mbox{\boldmath$\xi$}}

\renewcommand\baselinestretch{1.1}

\begin{document}
\rm

\begin{center}
{\sf VISCOSITY-STRATIFIED FLOW IN A HELE-SHAW CELL} \\
\vspace{2mm}

A.\,A. Chesnokov$^{1,2}$, V.\,Yu. Liapidevskii$^{1,2}$ \\[2mm]

${^1}$Novosibirsk State University, \\
Pirogova Str. 2, Novosibirsk, 630090, Russia \\[2mm]
${^2}$Lavrentyev Institute of Hydrodynamics SB RAS, \\
Lavrentyev Ave. 15, Novosibirsk, 630090, Russia \\ 
e-mails: chesnokov@hydro.nsc.ru, liapid@hydro.nsc.ru
\end{center}

\begin{abstract}
A hierarchy of mathematical models describing viscosity-stratified flow in a Hele-Shaw cell is constructed. Numerical modelling of jet flow and development of viscous fingers with the influence of inertia and friction is carried out. One-dimensional multi-layer flows are studied. In the framework of three-layer flow the interpretation of the Saffman--Taylor instability is given. A kinematic-wave model of viscous fingering taking into account friction between the fluid layers is proposed. Comparison with calculations on the basis of two-dimensional equations shows that this model allows to determine the velocity of propagation and the thickness of the viscous fingers. 
\end{abstract}


\section{Introduction} 
A displacement process involving two fluids is often unstable when the displacing fluid has larger viscosity than the displaced one. The resulting instability developing at the interface between two fluids is known as viscous fingering \cite{SaffmanTaylor, Homsy}. This instability has received much attention as an archetype of pattern-formation problems and as a limiting factor in the recovery of crude oil. Classical mathematical model describing a Newtonian flow displacement in a Hele-Shaw cell and development of the Saffman--Taylor instability consists of the continuity equation, Darcy's law and a convection-diffusion equation for the concentration of the displacing fluid \cite{TanHomsy, Azaiez}. The inertia of fluid may be important for high finger velocities. This leads to the necessity to use more complex nonlinear equations of fluid motion \cite{Gondret}. In the framework of these models instability caused by different velocities of layers movement can be considered. There are several theoretical and experimental studies on the role of inertia in immiscible \cite{Chevalier, DiasMiranda} and miscible \cite{YuanAzaiez} displacements. The results reveal that inertia tends to damping viscous fingering. In recent publications \cite{Housseiny, Pihler} the effects of the thickness variation of a Hele-Shaw cell and elasticity of the walls on the process of viscous fingering have been studied. Different types of instability in viscosity-stratified flow have been discussed in \cite{Sahu}. 

A number of theoretical, numerical, and experimental works devoted to understanding of various aspects of the instabilities is available in the literature \cite{Manickam, Tanveer, Smirnov}. Accurate numerical solution is costly at high Peclet numbers and it is difficult to reproduce the detailed fingering pattern. 
Numerical simulations \cite{Zimmerman, Yortsos02} show that it is simpler to describe the concentration of solute averaged across the fingers. In such cases, the mixing zone is an important feature to determine the extent of mixing. Despite the considerable work done, the spreading and growth of the mixing zone is an important question that still remains unresolved. Several empirical models are available for the evaluation of mixing zones in unstable, miscible displacements. Two empirical models have been suggested by Koval \cite{Koval} and Todd and Longstaff \cite{Todd} to give a basis for computation of miscible displacement. Both models suffer from adoption of empiricism in which the principal parameters involved have little or indirect physical significance. Further development of averaged models of fingers formation is represented in \cite{Fayers, Yortsos06, Booth}. All these models are based on the hypothesis of pressure equalization in the transverse direction to the main flow, as well as an empirical information about the displacing and displaced fluids distribution in the region of intensive viscous fingering. Let us recall that 2D numerical solution with high resolution is hard to construct, that is why 1D models play an important role in some cases. For instance, these models are very useful in calculating of fracturing when it is necessary to solve the equations of the crack opening and the fluid motion in the fracture simultaneously \cite{Adachi}. 

The aim of the present paper is to derive a hierarchy of mathematical models describing viscosity-stratified flow and spreading and growth of the mixing zone in a Hele-Shaw cell. In Section 2 we propose 2D nonlinear hyperbolic system of balance laws. In contrast with widespread model of flow displacement in a Hele-Shaw cell we apply nonlinear momentum equations and take into account compressibility of the fluid. At the same time diffusion coefficient is neglected that corresponds to the large Peclet number limit. As we show in Section 3 by numerical calculations, this model is suitable for describing of jet flow and propagation of viscous fingers in a Hele-Shaw cell. We also point out that for the process of unidirectional displacement the pressure variation in the transverse direction is small. This observation makes it possible to use long-wave approximation and construct a class of layered flows described by a system of one-dimensional evolution equations. Based on the various simplifications of the momentum equation (linearisation, lubrication theory) a hierarchy of 1D mathematical models is constructed in Section 4. The equations of a three-layer flow are studied and numerical computations of the formation of viscous fingers are performed in Section 5. We show that three-layer stationary flow is correctly described in the framework of simplified model. Nevertheless the growth rate of viscous fingers is significantly higher than it is observed experimentally. In section 6 we propose 1D kinematic-wave model of viscosity-stratified flow taking into account friction between the fluid layers. The velocity of propagation and the thickness of the viscous finger in the framework of this kinematic-wave model coincide with the corresponding calculations on the basis of the 2D equations. It gives possibility to predict the parameters of viscous fingers without time-consuming calculations. We also show that the proposed 1D model is in good agreement with the well-known Koval model. 

\begin{figure}[t]
\begin{center}
\resizebox{0.65\textwidth}{!}{\includegraphics{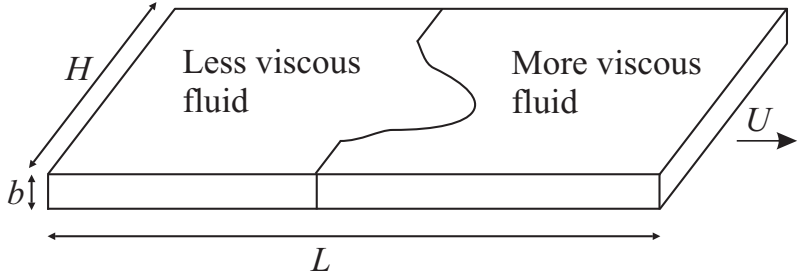}}\\[0pt]
{\caption{Hele-Shaw cell geometry.} \label{fig:fig_1}} 
\end{center}
\end{figure}

\section{Mathematical model} 
A Newtonian weakly-compressible flow displacement in a Hele-Shaw cell (the area between two parallel plates separated by a small gap of constant thickness $b$ in the $z$ direction, see Fig.~\ref{fig:fig_1}) is described by the equations
\begin{equation}\label{eq:N-St}
 \begin{array}{l}\displaystyle
  (\rho \bv)_t+{\rm div}(\rho\bv\otimes\bv-\bP)=0, \quad 
  \rho_t+{\rm div}(\rho\bv)=0, \quad \bv\big|_{z=\pm b/2}=0.
 \end{array}
\end{equation}
Here $\bv=(u,v,w)$ is the fluid velocity, $\rho$ is the density, and $\bP$ is the stress tensor. This tensor can be taken in the form $\bP=-\big(p+\frac{2}{3}\mu\,{\rm div}\,\bv\big)\bI+ \big(\nabla\bv+ (\nabla\bv)^*\big) \mu$, where $p$ is the pressure, and $\mu$ is the viscosity. To describe the process of displacement involving two fluids of different constant viscosities we should take into account that viscosity $\mu$ depends on the concentration of solvent $c$. This function is scaled such that it is equal to unity in the displaced fluid ($\mu=\mu_2$) and zero in the displacing one ($\mu=\mu_1$). Following~\cite{Azaiez} we assume a monotonic relationship between the viscosity and the concentration in the form $\mu(c)=\mu_1^{1-c}\mu_2^c$. The concentration $c$ satisfies to the transport equation (diffusion is neglected that corresponds to the large Peclet number limit)
\begin{equation}\label{eq:indecator} 
 c_t+\bv\cdot\nabla c=0. 
\end{equation}

We assume now that the velocity field can be represented as
\begin{equation}\label{eq:vel-3D} u=\frac{3}{2}\Big(1-\Big(\frac{2z}{b}\Big)^2\Big)u'(t,x,y), \quad 
   v=\frac{3}{2}\Big(1-\Big(\frac{2z}{b}\Big)^2\Big)v'(t,x,y), \quad w=0. 
\end{equation}
It provides the fulfilment of no-slip conditions on the cell walls $z=\pm b/2$.    
We also suppose that the functions $p$, $\rho$, and $c$ do not depend on $z$. Let us note that in the calculation of ${\rm div}\,\bP$ the following terms $(\mu\bv_x)_x$, $(\mu\bv_x)_y$, and $(\mu\bv_y)_y$ can be omitted as they are negligible compared to the derivatives with respect to $z$. Further, averaging Eqs.~\eqref{eq:N-St} and \eqref{eq:indecator} through the gap we obtain (primes are omitted)
\begin{equation}\label{eq:compr-HS}
 \begin{array}{l}\displaystyle
  (\rho u)_t+(\beta\rho u^2+p)_x+(\beta\rho uv)_y=-\mu u, \\[2mm]\displaystyle
  (\rho v)_t+(\beta\rho uv)_x+(\beta\rho v^2+p)_y=-\mu v, \\[2mm]\displaystyle
  \rho_t+(u\rho)_x+(v\rho)_y=0, \quad 
  (c\rho)_t+(uc\rho)_x+(vc\rho)_y=0. 
 \end{array}
\end{equation}
Here and below $\mu$ denotes the viscosity of the fluid divided by the permeability $b^2/12$ (further in the text we will call it simply ``viscosity''); coefficient $\beta$ is equal $6/5$ (this factor comes from integration of $\bv\otimes\bv$ in the form \eqref{eq:vel-3D} with respect to $z$ \cite{Gondret, DiasMiranda}). 

In order to close model \eqref{eq:compr-HS} we should specify either the equation of state $p=p(\rho)$ (barotropic fluid) or dependence $\rho=\rho(c)$ (incompressible fluid). A weak compressibility of the fluid given by the equation of state $p=p(\rho)$ provides hyperbolicity of the model. This dependence can be considered as a regularization of equations describing the flow of an incompressible fluid in a Hele-Shaw cell. On the other hand in applications the property of hyperbolicity of equations can be related to the presence of gas cavities in the porous medium as well as with the elasticity of the channel walls (for instance in PKN model \cite{Adachi} the pressure $p$ depends on the channel thickness $b$). In any case if the condition $u^2+v^2 \ll p'(\rho)$ holds then results weakly depend on the choice of $p=p(\rho)$. Therefore for the numerical simulation of 2D flows we assume 
\begin{equation}\label{eq:p-rho} 
  p(\rho)=a^2\rho^2/2 \quad (a^2=c_0^2/\rho_0) 
\end{equation}
where the constants $\rho_0$ and $c_0$ specify characteristic density and speed of sound in the fluid. 

To find the characteristics of system \eqref{eq:compr-HS}, \eqref{eq:p-rho} we write it in the vector form 
\[ \bU_t+\bA\bU_x+\bB\bU_y=\bF \]
where $\bU=(u,v,\rho,c)^{\rm T}$ is the vector of dependent variables; $\bF=(-\mu u/\rho,-\mu v/\rho,0,0)^{\rm T}$ is the right-hand side; $\bA$ and $\bB$ are $4\times 4$ matrices. Let $\bxi=(\xi_1,\xi_2,\xi_3)$ be the normal vector to the characteristics; $\bI$ is the identity matrix. Then the characteristic matrix $\bC=\xi_1\bI+\xi_2\bA+\xi_3\bB$ of system \eqref{eq:compr-HS}, \eqref{eq:p-rho} has the form
\[ \bC=
 \begin{pmatrix}
\chi_1+(\beta-1)u\xi_2 & (\beta-1)u\xi_3 & \big((\beta-1)(\chi_2-\xi_1)u+p'(\rho)\xi_2\big)\rho^{-1} & 0 \\[1mm]
(\beta-1)v\xi_2 & \chi_1+(\beta-1)v\xi_3 & \big((\beta-1)(\chi_2-\xi_1)v+p'(\rho)\xi_3\big)\rho^{-1} & 0 \\[1mm]
\rho\xi_2 & \rho\xi_3 & \chi_2 & 0 \\[1mm]
0 & 0 & 0 & \chi_2
 \end{pmatrix} \,. \]
Here $\chi_1=\xi_1+\beta u\xi_2+\beta v\xi_3$, $\chi_2=\xi_1+u\xi_2+v\xi_3$. A simple but cumbersome calculation yields the following expression for ${\rm det}\,\bC(\bxi)$:
\[ {\rm det}\,\bC(\bxi)=\big((\xi_1^2+2\beta(u\xi_2+v\xi_3)\xi_1+ \beta(u\xi_2+v\xi_3)^2)-
   (\xi^2+\eta^2)p'(\rho)\big)\chi_1\chi_2. \]
We specify the characteristic surface by the equation $W(t,x,y)=0$. Then to obtain the differential equations of the characteristics we should replace the vector $(\xi_1,\xi_2,\xi_3)$ in previous equation by the vector $(W_t,W_x,W_y)$ and equate to zero ${\rm det}\, \bC$. As a result we obtain two families of contact characteristics 
\[ W_t+uW_x+vW_y=0, \quad W_t+\beta uW_x+\beta vW_y=0 \] 
and two additional characteristic families 
\[ W_t+\beta uW_x+\beta vW_y=\pm\sqrt{\beta(\beta-1)(uW_x+vW_y)^2+(W_x^2+W_y^2)p'(\rho)}\,. \] 
If the inequalities $\beta\geq 1$ and $p'(\rho)>0$ hold this system of equations is hyperbolic.
Note that in the case of $\beta=1$, $\mu=0$, and $c={\rm const}$ system \eqref{eq:compr-HS}, \eqref{eq:p-rho} coincides with the well-known shallow water equations. 

\section{Modelling of viscous fingering and jet flows} 
Below we present the results of numerical calculations of the viscous fingers and jet streams on the basis of hyperbolic model \eqref{eq:compr-HS}, \eqref{eq:p-rho}. Originally mathematical description of the Saffman--Taylor instability was given in the framework of the Darcy's law and the mass conservation equation \cite{SaffmanTaylor, TanHomsy}. The inertia of the fluid may be significant for high finger velocities. In \cite{Chevalier} simulation of viscous fingers was performed using non-linear equations of an incompressible fluid. We show that Eqs.~\eqref{eq:compr-HS}, \eqref{eq:p-rho} taking into account the forces of inertia and compressibility of the fluid could be also used for description of this instability.

To solve differential balance laws \eqref{eq:compr-HS}, \eqref{eq:p-rho} numerically one can apply methods based on various modifications of Godunov's scheme. In this work we implement the robust and stable Nessyahu--Tadmor second-order central scheme \cite{NT90}. In every test we assume that on the boundaries $y=0$ and $y=H$ the impermeability condition $v=0$ is fulfilled. The size of the computational domain is $L=100$, $H=50$; the resolution of the problem on the $x$ and $y$ axes are 300 and 150 nodes correspondingly (uniform grid). We assume that $\rho_0=1$, $c_0=150$ and $\beta=6/5$ (for the third test $\beta=1$). The values of the variables are considered as dimensionless. Below we present calculations showing the possibility of modelling the evolution of perturbations caused by the Kelvin--Helmholtz and/or Saffman--Taylor instabilities on the basis of the hyperbolic model \eqref{eq:compr-HS}, \eqref{eq:p-rho}. 

\subsection{Test 1. Jet flow}
\begin{figure}[t]
\begin{center}
\resizebox{0.49\textwidth}{!}{\includegraphics{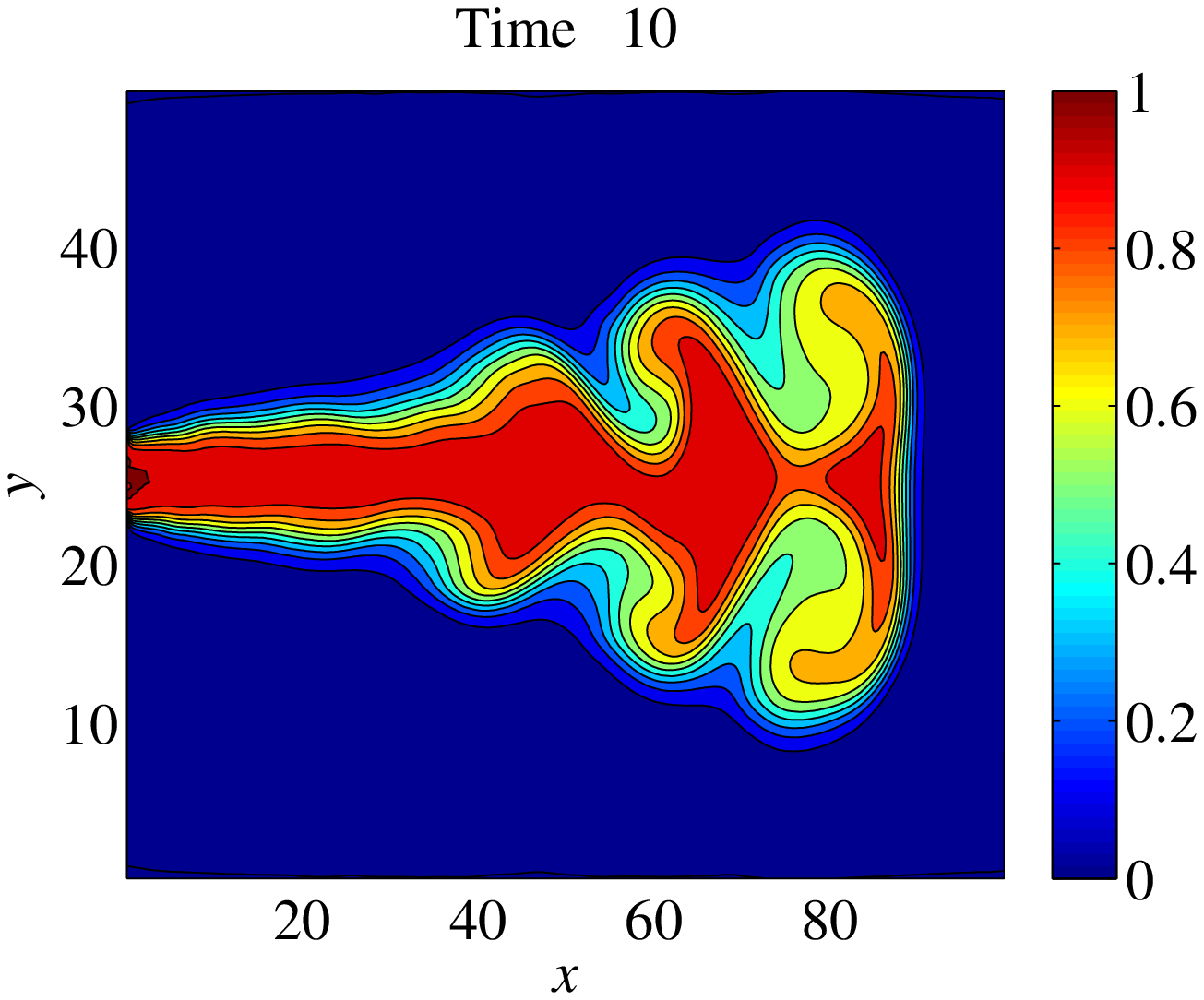}}\hfill
\resizebox{0.49\textwidth}{!}{\includegraphics{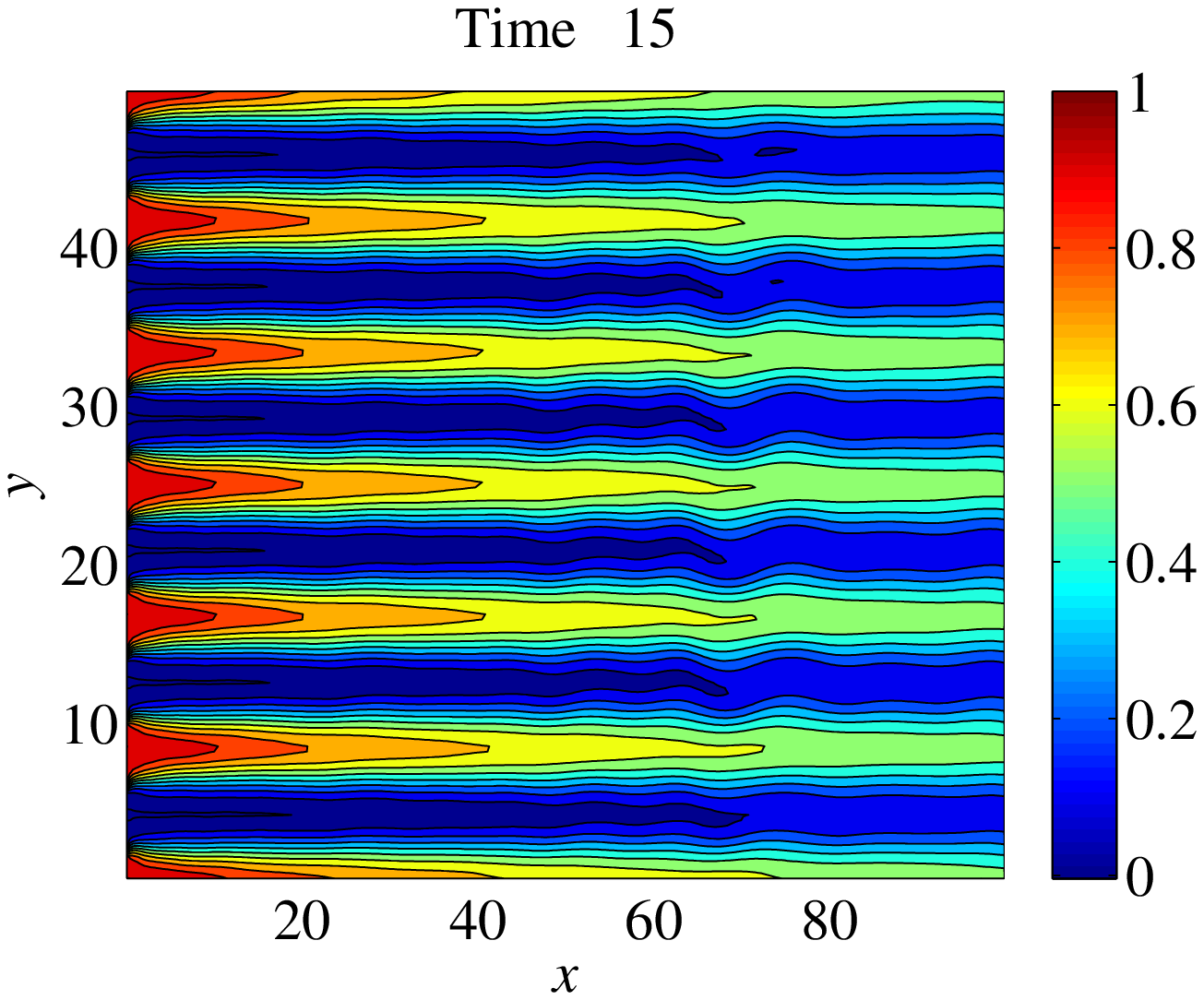}}\\[0pt]
\parbox{0.47\textwidth}{\caption{Kelvin--Helmholtz instability in jet flow of  viscosity-stratified fluid injected at $x=0$ ($\mu_1=0.1$, $U_1=4$, and $\mu_2=0.4$, $U_2=24$).} \label{fig:fig_2}} \hfill
\parbox{0.47\textwidth}{\caption{Modelling of multi-layer jet flow. Fluid of viscosity $\mu_1=1$ and $\mu_2=3$ (for odd and even layers) is injected at $x=0$ with velocities $U_1=21$, and $U_2=7$.} \label{fig:fig_3}}
\end{center}
\end{figure}

Let at the initial time $t=0$ the Hele-Shaw cell be occupied by a quiescent fluid having density $\rho=1$ and viscosity $\mu_1=0.1$. Through the left central cross-section of width $H/10$ fluid of viscosity $\mu_2=0.4$ is injected with velocity $U_2=24$; through the rest part of the left boundary fluid of viscosity $\mu_1=0.1$ is entered with velocity $U_1=4$. On the right boundary of the domain the condition of constant pressure is valid. For more intensive development of the perturbations at each time step we slightly disturb boundary conditions at $x=0$. Namely, the cross section, through which fluid ``2'' is injected, is randomly shifted up/down from its initial position on fixed distance $\Delta y$ which is equal to the grid spacing with some positive integer factor $k$. We call it ``random shake'' and take here $k=1$. The function $c$ for values of the concentration is presented in Fig.~\ref{fig:fig_2} at $t=10$. Vortices are formed at the interface of the layers due to Kelvin--Helmholtz instability. Increase of the both values $\mu_1$ and $\mu_2$ suppresses this instability. 

\subsection{Test 2. Viscosity-stratified flow}
Viscosity-stratified multi-layer flow is shown in Fig.~\ref{fig:fig_3} at $t=15$. Initially the flow region is filled by a quiescent fluid ($\mu_1=1$, $\rho=1$). Liquid is pumped through the left boundary divided into layers of height $H/12$; velocity and viscosity for odd and even layers are $U_2=7$, $\mu_2=3$ and $U_1=21$, $\mu_1=1$, correspondingly. As in the previous example ``random shake'' of the jets is used. The results of the calculations show that multi-layer flow without mixing is realized for a wide range of parameters (Kelvin--Helmholtz instability at the interfaces between the layers occurs if viscosity decreases more than in five times). We note that the condition on the left boundary $U\mu={\rm const}$ corresponds to a class of exact solutions of equations \eqref{eq:compr-HS} for an incompressible fluid: $u=U(y)$, $v=0$, $\rho={\rm const}$, $p=-\alpha x$, $\mu=\alpha/U(y)$. The stability analysis of this class of flows was carried out in~\cite{ChSt15}.

\subsection{Test 3. Viscous fingering}
The following example illustrates the formation of viscous fingers. Let the displacing phase injected at a constant velocity $U$ be referred to with index 1 and the displaced one with index 2. 
At $t=0$ fluid ``1'' is located in the domain $x<x_0=L/2$; more viscous fluid ``2'' --- in the domain $x>x_0$. For convenience we use the coordinate system moving with velocity $U$ and assume $\beta=1$ (with $\beta=6/5$ results are similar). Let us perturb the initial interface $x=x_0$ as follows: $x=x_0+\Delta y\,\cos(5\pi y/H)$ (here $\Delta y=2/3$). We note that the physical effect of the instability can be obtained numerically if the initial perturbation is not less than the grid resolution. At $t=0$ we choose piece-wise linear pressure distribution ($p_x=-\mu_1 U$ for $x<x_0$ and $p_x=-\mu_2 U$ for $x>x_0$, on the right boundary $p$ is equal to $c_0^2\rho_0/2$). Initial density of the fluid is determined using formula~\eqref{eq:p-rho}. At the boundaries the impermeability condition is fulfilled. 

\begin{figure}[t]
\begin{center}
\resizebox{0.49\textwidth}{!}{\includegraphics{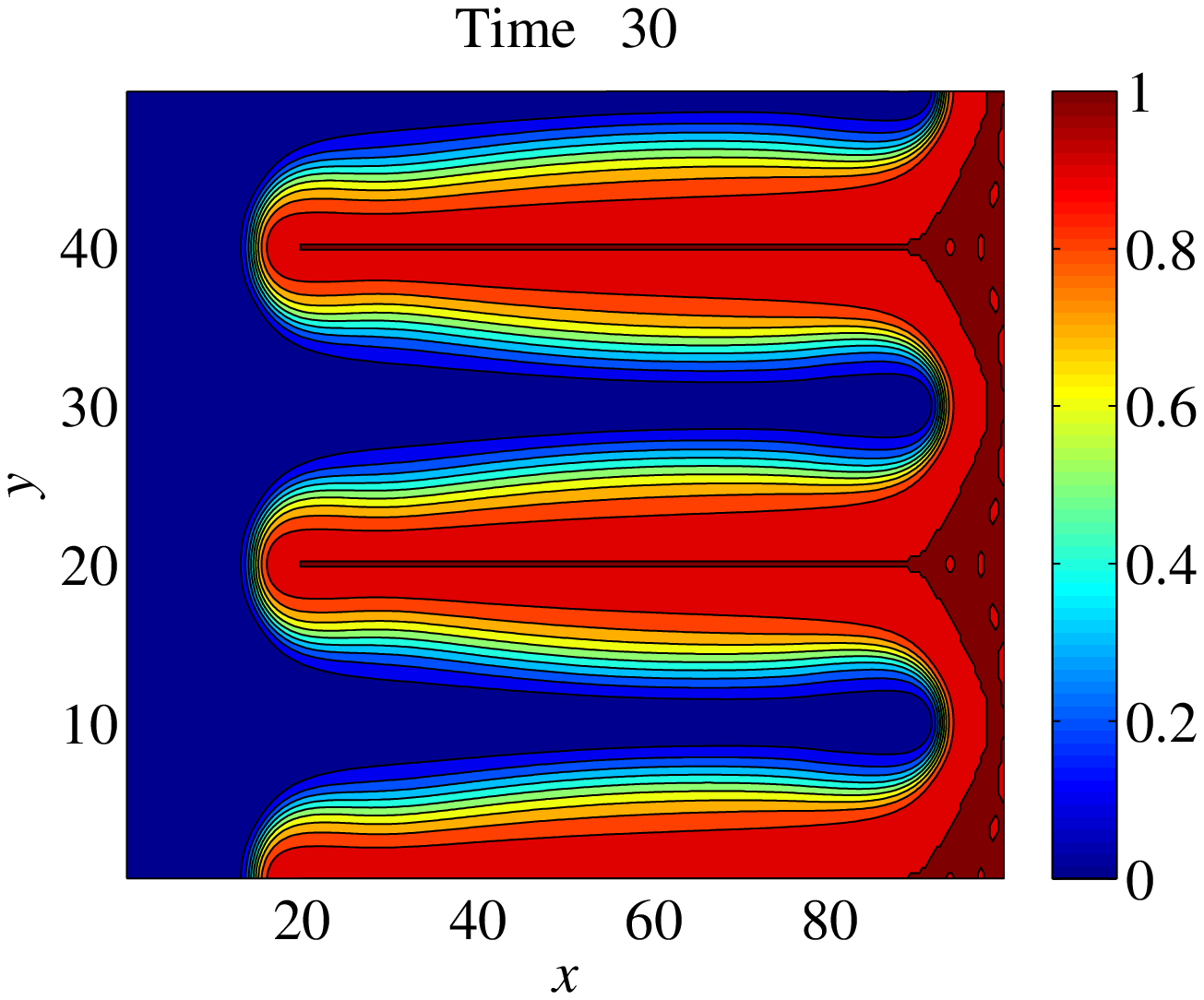}}\hfill
\resizebox{0.49\textwidth}{!}{\includegraphics{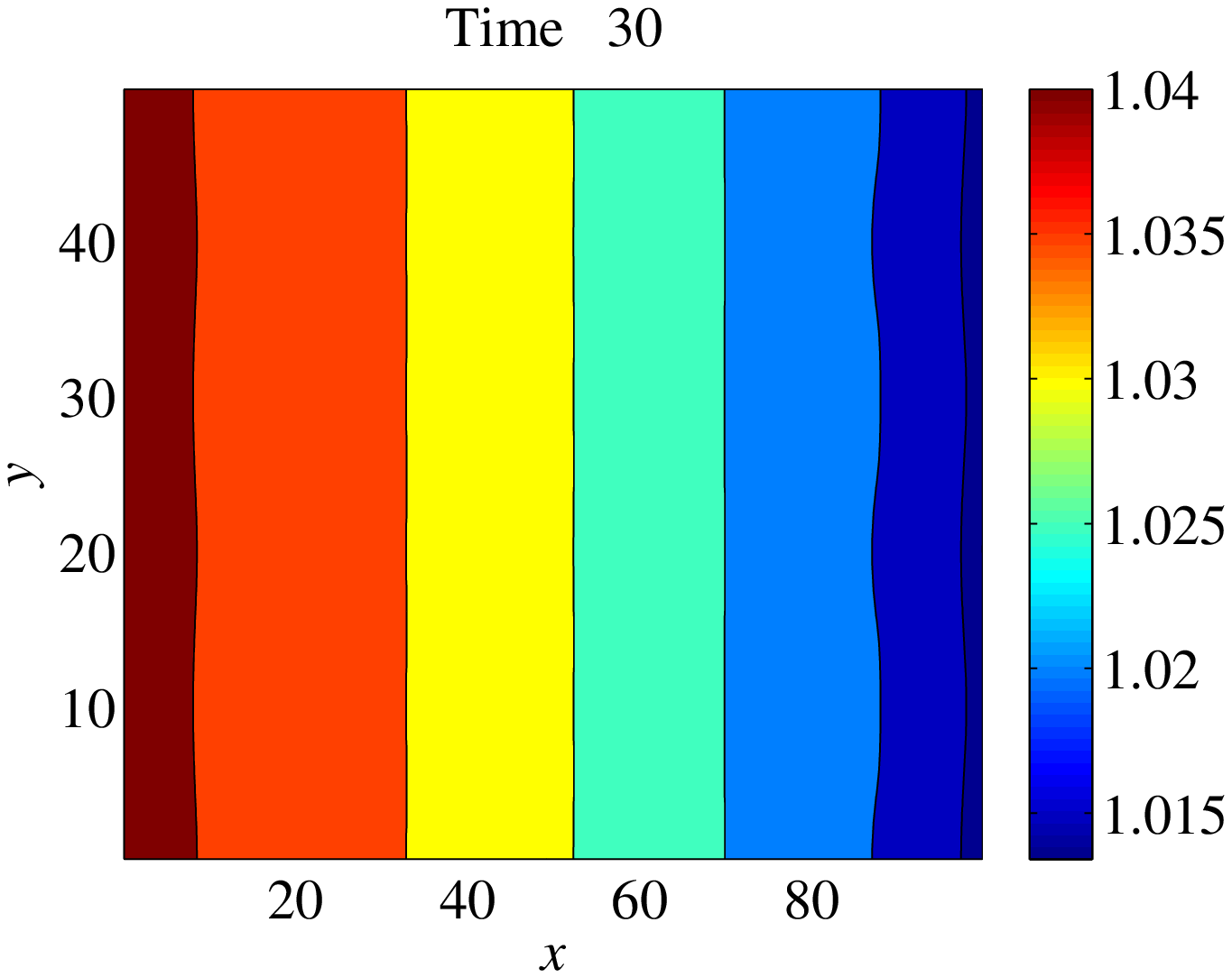}}\\[0pt]
{\caption{Formation of viscous fingers in weakly-compressible Hele-Shaw flow for $U=3$, $\mu_1=1$, and $\mu_2=5$: the concentration $c$ (on the left) and the density $\rho$ (on the right).} \label{fig:fig_4}} 
\end{center}
\end{figure}

The calculations are performed for the following parameters: $U=3$, $\mu_1=1$, and $\mu_2=5$. In the  evolution process of the flow viscous fingers are formed (Fig.~\ref{fig:fig_4}, left). The number of fingers is determined by the initial perturbation. Displacing fluid penetrates more rapidly into displaced one (fingers are not symmetrical with respect to the initial interface). Fig.~\ref{fig:fig_4} (right) shows the distribution of the density at $t=30$. Calculation with better resolution leads to the same result. As we can see the density changes less than $0.5\%$ in comparison with the initial one (to reduce the compressibility we should increase the speed of sound $c_0$ but this slows down the calculations since the time step is determined by the Courant number). Note that the density (pressure) varies slightly with respect to $y$. This allows one to use approximate model, where the second momentum equation is replaced by $p_y=0$. 

\section{Layered flows} 
We consider an incompressible ($\rho=1$) Hele-Shaw flow in a two-dimensional domain of rectangular geometry (dimensions $L$ and $H$, respectively) governed by Eqs.~\eqref{eq:compr-HS}. It is assumed that the flow is essentially parallel and the pressure gradient in the flow direction $x$ being independent of the transverse coordinate $y$ to leading-order in the variable $\varepsilon=H/L$. This regime is termed as the state of transverse flow equilibrium \cite{Yortsos06}. Such flows can be also considered in the framework of long wave approximation \cite{LT00, ChL11}. Let us perform the following scaling in Eqs.~\eqref{eq:compr-HS} 
\[ t\to \varepsilon^{-1}t, \quad x\to \varepsilon^{-1}x, 
   \quad v\to \varepsilon v, \quad \mu\to \varepsilon \mu. \]
Then we neglect terms of order $\varepsilon^2 \ll 1$. As a result we obtain the approximate model
\begin{equation}\label{eq:model-LW}
 \begin{array}{l}\displaystyle
   u_t+\beta uu_x+\beta vu_y+p_x=-\mu u, \quad p_y=0, \\[2mm]\displaystyle
   u_x+v_y=0, \quad c_t+uc_x+vc_y=0, \\[2mm]\displaystyle
   v\big|_{y=0}=0, \quad v\big|_{y=H}=0,
 \end{array} 
\end{equation}
where the pressure does not depend on the variable $y$. We also suppose that on the boundaries $y=0$ and $y=H$ the impermeability condition is fulfilled. 

Let us consider the class of viscosity-stratified flows 
\[ u=u_i(t,x), \quad c=c_i={\rm const}, \quad y\in (y_{i-1},y_i) \]
$(0=y_0<y_1(t,x)<...<y_N=H)$. In this case Eqs.~\eqref{eq:model-LW} take the form
\begin{equation}\label{eq:layers} 
  \begin{array}{l}\displaystyle
    u_{it}+\beta u_iu_{ix}+p_x=-\mu_i u_i, \quad h_{it}+(u_i h_i)_x=0, 
    \quad (i=1,...,N) \\[2mm]\displaystyle
    \sum\limits_{i=1}^N h_i=H, \quad \sum\limits_{i=1}^N u_i h_i=Q.
  \end{array} 
\end{equation}
Here $h_i(t,x)=y_i(t,x)-y_{i-1}(t,x)$ is the depth of $i-$th liquid layer of viscosity $\mu_i$  having velocity $u_i(t,x)$; and $Q$ is the total flow rate through the cell. Upon derivation of Eqs.~\eqref{eq:layers} the kinematic condition at the layers interface is used. 

Introducing new unknown variables $s_i=u_i-u_N$ allows to transform Eqs.~\eqref{eq:layers} to the evolution system of $2(N-1)$ equations 
\[ \begin{array}{l}\displaystyle 
    s_{it}+\beta((s_i/2+u_N)s_i)_x=(\mu_N-\mu_i)u_N-\mu_i s_i, \\[2mm]\displaystyle 
    h_{it}+((s_i+u_N)h_i)_x=0, \quad\quad (i=1,...,N-1)
   \end{array} \]
where 
\[ h_N=1-\sum\limits_{i=1}^{N-1} h_i, \quad 
   u_N=\frac{1}{H}\Big(Q-\sum\limits_{i=1}^{N-1} s_i h_i\Big). \]
Further we assume that $Q={\rm const}$. 

We also use the following simplified versions of governing Eqs.~\eqref{eq:layers}. The first one consists in the linearisation of the momentum equations: 
\[ u_{it}+\beta U u_{ix}+p_x=-\mu_i u_i \quad (i=1,...,N) \]
(here $U=Q/H$ is the average velocity). The second simplification is based on the Darcy law:
\[ p_x=-\mu_i u_i \quad (i=1,...,N). \]
The remaining equations of system \eqref{eq:layers} do not vary. 

In some cases it is convenient to use a moving coordinate system $x'=x-Ut$, $u'_i=u_i-U$. Then  Eqs.~\eqref{eq:layers} take the form (primes are omitted) 
\begin{equation}\label{eq:layers-mov} 
  \begin{array}{l}\displaystyle
    u_{it}+(\beta u_i+(\beta-1)U)u_{ix}+p_x=-\mu_i(u_i+U), \\[2mm]\displaystyle 
    h_{it}+(u_i h_i)_x=0, \quad \sum\limits_{i=1}^N h_i=H, \quad \sum\limits_{i=1}^N u_i h_i=0.
  \end{array} 
\end{equation}
Further we show that in the framework of three-layer and two-layer regimes of flow it is possible to give an interpretation of the Saffman--Taylor instability as well as to describe the initial stage of viscous fingering. 

\section{Three-layer flow} 
We introduce the following notation for the layer velocities and depths 
\[ u=u_1, \ v=u_2, \ w=u_3; \ h=h_1, \ \eta=h_2, \ \zeta=h_3. \]
We also assume that $H=1$, $Q=1$, $\mu_1=\mu_3=1$, and $\mu_2=\mu\neq 1$. 

\subsection{Stationary solutions} 
\begin{figure}[t]
\begin{center}
\resizebox{0.44\textwidth}{!}{\includegraphics{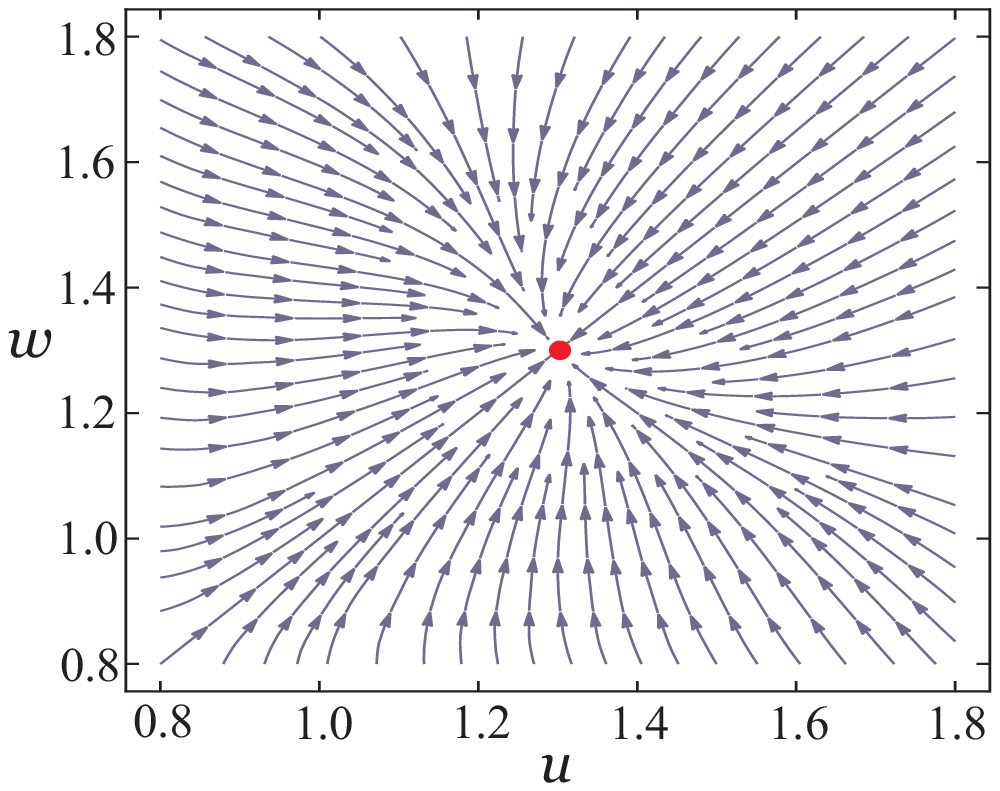}}\hfill
\resizebox{0.49\textwidth}{!}{\includegraphics{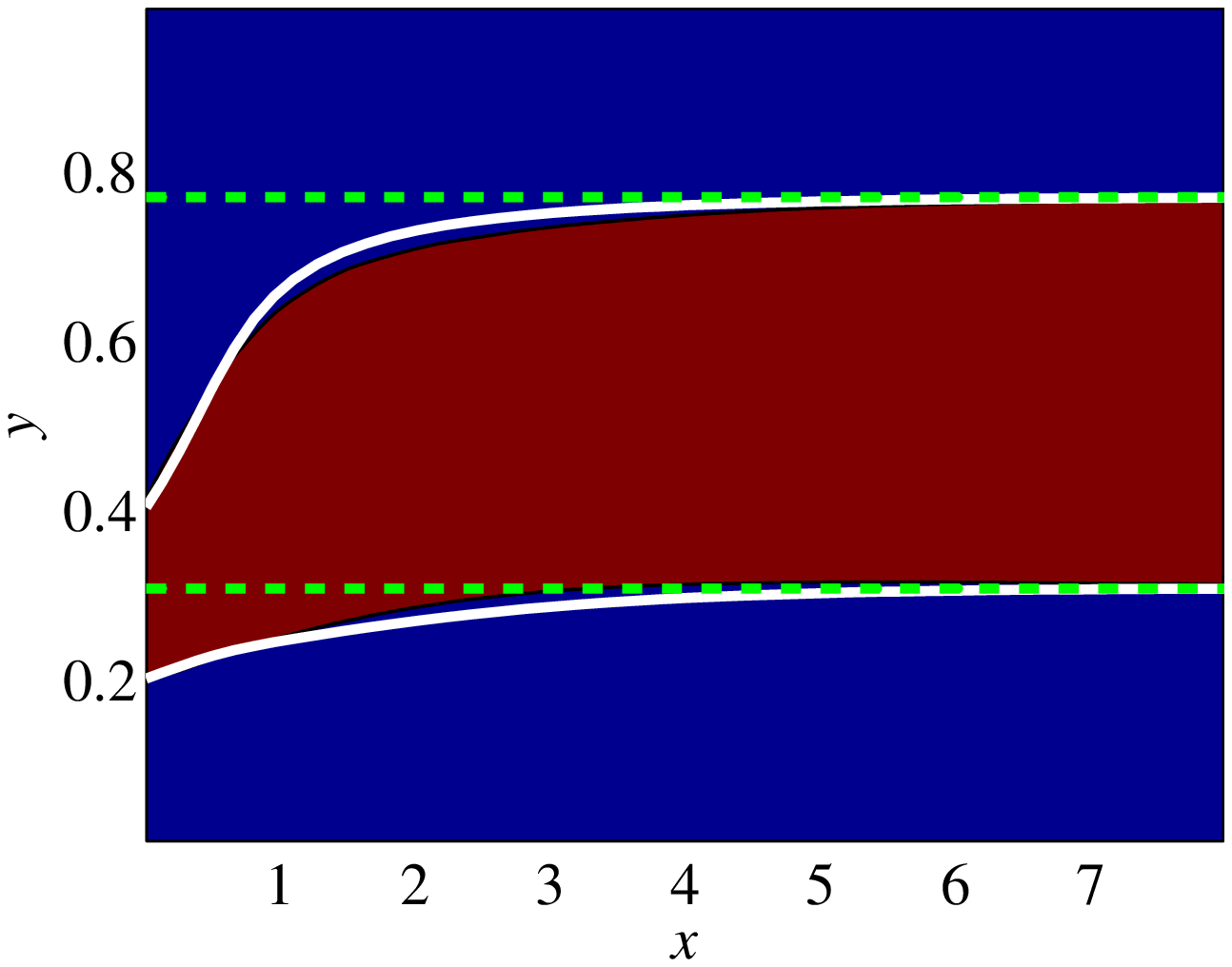}}\\[0pt]
\parbox{0.48\textwidth}{\caption{The integral curves of ODE \eqref{eq:st-sol} in the phase plane (a stable node) obtained for $Q_1=0.4$, $Q_2=Q_3=0.3$, and $\mu=2$.} \label{fig:fig_5}} \hfill
\parbox{0.48\textwidth}{\caption{Comparison of the results for 2D Eqs.~\eqref{eq:compr-HS}, \eqref{eq:p-rho} (concentration $c$ is shown in two colours) and Eqs.~\eqref{eq:st-sol} for  three-layer flow (solid white lines).} \label{fig:fig_6}}
\end{center}
\end{figure}

Here we construct a steady-state solution of model \eqref{eq:layers} for three-layer flow. Integration of the equations of conservation of mass in system \eqref{eq:layers} allows to express the depths of the layers
\begin{equation}\label{eq:h-eta} 
  h=Q_1/u, \quad \eta=Q_2/v, \quad \zeta=Q_3/w.
\end{equation}
Here $Q_i$ is the flow rate in the $i$-th layer ($Q_1+Q_2+Q_3=1$). Due to the unit depth we obtain the velocity in the intermediate layer 
\[ v=\varphi(u,w)=\frac{Q_2}{\Delta}, \quad \Delta=1-\frac{Q_1}{u}-\frac{Q_3}{w}\,. \]
Eliminating pressure $p$ from the equations
\[ \beta uu'+p'=-u, \quad \beta vv'+p'=-\mu v, \quad \beta ww'+p'=-w \]
(here the prime denotes the derivative with respect to $x$) reduces the problem to the solution of the autonomous system of ordinary differential equations
\begin{equation}\label{eq:st-sol}  
  \frac{du}{dx}=\frac{(u-\mu\varphi)w-(u-w)\varphi\varphi_w}
  {((\varphi\varphi_u-u)w+u\varphi\varphi_w)\beta}, \quad 
  \frac{dw}{dx}=\frac{(u-\mu\varphi)u+(u-w)(\varphi\varphi_u-u)}
  {((\varphi\varphi_u-u)w+u\varphi\varphi_w)\beta}\,.
\end{equation}
A fixed point of system \eqref{eq:st-sol} is determined from the relations $u=w=\mu\varphi$: 
\[ u_*=w_*=1+(\mu-1)Q_2. \]
Linearisation of Eqs.~\eqref{eq:st-sol} on the solution $u=u_*$, $w=w_*$ and computation of the eigenvalues of the corresponding matrix show that the fixed point is a stable node. The integral curves in the phase plane $(u,w)$ in the neighbourhood of the fixed point are shown in Fig.~\ref{fig:fig_5} for $Q_1=0.4$, $Q_2=Q_3=0.3$, $\mu=2$, and $\beta=6/5$.

Fig.~\ref{fig:fig_6} shows a comparison of the numerical results obtained on the basis of two-dimensional hyperbolic equations~\eqref{eq:compr-HS}, \eqref{eq:p-rho} and multilayer model~\eqref{eq:layers} reduced to dynamical system~\eqref{eq:st-sol} in the case of three-layer stationary flow. Solid white lines indicate the layers depths $y=h(x)$ and $y=h(x)+\eta(x)$ obtained by solving equation~\eqref{eq:st-sol} and using relations~\eqref{eq:h-eta}; dotted lines correspond to the fixed point ($h_*=Q_1/u_*$, $h_*+\eta_*=1-Q_3/w_*$). Here we take the following values of layers depths $h_0=\eta_0=0.2$, $\zeta_0=0.6$ at $x=0$. As before we choose $Q_1=0.4$, $Q_2=Q_3=0.3$, $\mu=2$, and $\beta=6/5$. The figure shows that the solution reaches an equilibrium state for $x>5$. 

To carry out the calculation on the basis of 2D equations \eqref{eq:compr-HS}, \eqref{eq:p-rho} the following initial data are used. The flow domain in the $y$-direction is divided into three layers of width $h_0$, $\eta_0$, and $\zeta_0$. The fluid of density $\rho=1$ at $t=0$ moves in these layers in the $x$-direction with constant velocities $u=Q_1/h_0$, $v=Q_2/\eta_0$, and $w=Q_3/\zeta_0$ respectively. We also suppose that $\mu=1$ in the layers of width $h_0$ and $\zeta_0$; in the middle layer of width $\eta_0$ we choose $\mu=2$. The same data are taken as the boundary conditions at $x=0$; on the right boundary ($x=8$) the condition of constant pressure is prescribed; on the walls $y=0$ and $y=1$ the condition of impermeability is fulfilled. The resolution of the problem on the $x$ and $y$ axes are 300 and 60 nodes correspondingly (uniform grid). In order to visualize the flow of fluids with different viscosities the concentration $c$ is used. This value is presented in Fig.~\ref{fig:fig_6} at $t=25$. More viscous fluid in the middle layer is shown in brown ($c>0.35$) and less viscous one is shown in blue ($c<0.35$). 

\subsection{Non-stationary solutions} 
Let us consider a three-layer flow governing by Eqs.~\eqref{eq:layers} wherein the momentum equations are replaced by linear Darcy laws $p_x=-\mu_i u_i$. Taking into account assumptions above and notations we have
\begin{equation}\label{eq:3l-represent} 
  u=w=\mu/d, \quad v=1/d, \quad d=(1-\mu)\eta+\mu. 
\end{equation}
In this case the depths of the layers $h$ and $\eta$ are found from the system of equations 
\begin{equation}\label{eq:3l-Darcy} 
  h_t+(u(\eta)h)_x=0, \quad \eta_t+(v(\eta)\eta)_x=0. 
\end{equation}
It is easy to check that this system is hyperbolic and its characteristic velocities are 
\[ \lambda_1=\eta v'(\eta)+v(\eta), \quad \lambda_2=u(\eta). \]
The first family of characteristics is genuinely nonlinear whereas the second one is linearly degenerate \cite{RYa78}. In terms of the Riemann invariants $\eta$ and $r=h/(1-\eta)$ Eqs.~\eqref{eq:3l-Darcy} take the form 
\[ \eta_t+\lambda_1(\eta)\eta_x=0, \quad r_t+\lambda_2(\eta)r_x=0. \]

\begin{figure}[t]
\begin{center}
\resizebox{.97\textwidth}{!}{\includegraphics{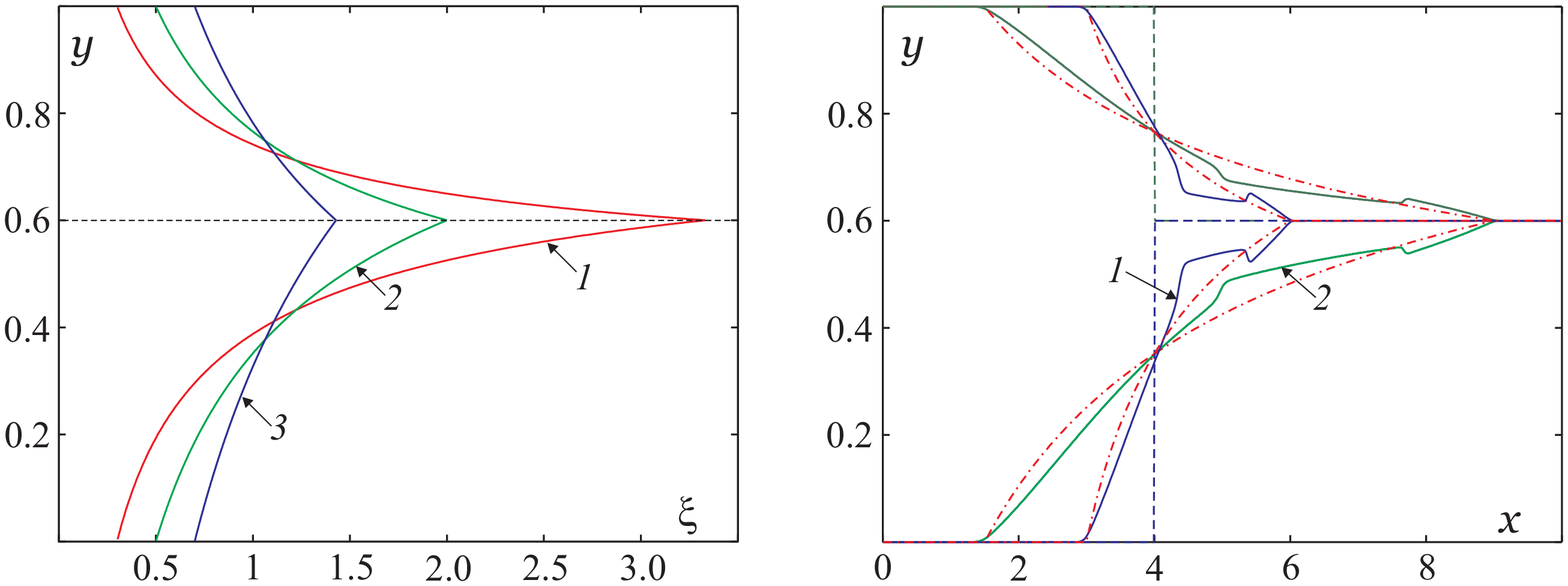}}\\[0pt]
\parbox{0.48\textwidth}{\caption{Layers thickness $y=h$ and $y=h+\eta$ obtained by self-similar solution \eqref{eq:3l-h-eta}: {\it 1} --- $\mu=0.3$, {\it 2} --- $\mu=0.5$, {\it 3} --- $\mu=0.7$.} \label{fig:fig_7}} \hfill
\parbox{0.48\textwidth}{\caption{Profiles of $y=h$ and $y=h+\eta$ (solution of \eqref{eq:3l-semilin} is given by solid curves, dot-dash corresponds to \eqref{eq:3l-h-eta}; dashed line presents the initial data): {\it 1} --- $t=2$, {\it 2} --- $t=5$.} \label{fig:fig_8}}
\end{center}
\end{figure}

In the case $0<\mu<1$ we construct a centred simple wave solution defined by the relations
\[ r=h_0={\rm const}, \quad \lambda_1(\eta)=\xi, \quad \xi=(x-x_0)/t \]
(note that the ansatz $\eta={\rm const}$ leads to a constant solution). The layer depths are 
\begin{equation}\label{eq:3l-h-eta} 
  \eta(\xi)=\frac{1}{1-\mu}\Big(\sqrt{\frac{\mu}{\xi}}-\mu\Big), \quad h(\xi)=(1-\eta)h_0 \quad 
  \Big(\mu<\xi<\frac{1}{\mu}\Big). 
\end{equation}
Formulae \eqref{eq:3l-h-eta} give the solution of Eqs.~\eqref{eq:3l-Darcy} with discontinuous initial data
\begin{equation}\label{eq:3l-Darcy-t0}  
  (h,\eta)\big|_{t=0} =
   \left\{
   \begin{array}{ll}
     (0, \ 1),   & \quad x<x_0 \\[2mm]
     (h_0, \ 0), & \quad x>x_0.
   \end{array}
  \right. 
\end{equation}
Profiles of a viscous finger $y=h$ and $y=h+\eta$ given by \eqref{eq:3l-h-eta} are shown in Fig.~\ref{fig:fig_7} for $h_0=0.6$ and various values of $\mu<1$.

Let us construct a solution of Cauchy problem \eqref{eq:3l-Darcy}, \eqref{eq:3l-Darcy-t0} for $\mu>1$. At initial time the velocities and the layers depths are $u^-=\mu$, \ $u^+=1$, \ $h^-=0$, \ $h^+=h_0$; \ $v^-=1$, \ $v^+=1/\mu$, \ $\eta^-=1$, \ $\eta^+=0$. It is easy to verify that these values satisfy the Hugoniot conditions 
\[ [(u-D)h]=0, \quad [(v-D)\eta]=0 \]
derived from Eqs.~\eqref{eq:3l-Darcy} as well as the stability conditions \cite{RYa78} if the shock front moves with average flow velocity $D=U=1$. 

It is interested to note that in the class of simple wave solutions ($r={\rm const}$) system~\eqref{eq:3l-Darcy} reduces to the na\"{i}ve Koval model \cite{Booth, Yortsos06}. In fact, taking into account representation \eqref{eq:3l-represent} the second equation in \eqref{eq:3l-Darcy} can be rewritten as 
\begin{equation}\label{eq:Koval}  
  \frac{\partial \bar{c}}{\partial t}+ \frac{\partial }{\partial x} 
  \bigg(\frac{M\bar{c}}{M\bar{c}+1-\bar{c}}\bigg) =0 
\end{equation}
where $\bar{c}$ stands for $\eta$ and $M=1/\mu$. It is known that the growth rate of viscous fingers in the framework of the na\"{i}ve Koval model is significantly higher than it is observed experimentally. 

To derive another model of a three-layer flow in a moving coordinate system we use linearisation  of momentum equations in \eqref{eq:layers-mov}
\[ u_{it}+\gamma Uu_{ix}+p_x=-\mu_i(u_i+U), \quad (\gamma=\beta-1). \] 
Eliminating the pressure $p$ leads to the system of evolution equations 
\begin{equation}\label{eq:3l-semilin}  
  \begin{array}{l}\displaystyle  
    s_{1t}+\gamma s_{1x}=-s_1, \\[2mm]\displaystyle 
    s_{2t}+\gamma s_{2x}=-\mu s_2+(1-\mu)(1-w), \\[2mm]\displaystyle
    h_t+(u h)_x=0, \quad \eta_t+(v\eta)_x=0
  \end{array} 
\end{equation}
where
\[ u=s_1+w, \quad v=s_2+w, \quad w=-hs_1-\eta s_2. \]

Let us rewrite Eqs.~\eqref{eq:3l-semilin} in the form $\bU_t+\bA\bU_x=\bF$. Here $\bU=(s_1, s_2, h, \eta)^{\rm T}$ is the unknown vector, $\bA$ is the Jacobi matrix, and $\bF$ is the right-hand part. The eigenvalues of matrix $\bA$ are
\[ \begin{array}{l}\displaystyle
     \lambda_1=\gamma s_1, \quad \lambda_2=\gamma s_2, \\[2mm]\displaystyle
     \lambda_{3,4}=2^{-1}\big((1-3h)s_1+(1-3\eta)s_2\pm\sqrt{m}\big) 
   \end{array} \]
where $ m=(s_2-(1-h)s_1)^2+\eta^2s_2^2+2((1+h)s_1-s_2)\eta s_2$. Conditions $0<h<1$ and $0<\eta<1$ provide hyperbolicity of system \eqref{eq:3l-semilin} since $m>0$. Introducing the parabolic with respect to $\tau=s_1/s_2$ function $f=m/s_2^2$ it is easy to check validation of inequalities $f''(\tau)>0$ and $f(\tau_*)>0$ (here $\tau_*$ is a minimum point of $f(\tau)$). It means that  $m>0$ and, consequently, characteristic velocities $\lambda_i$ are real. 

Further we construct the numerical solution of Eqs.~\eqref{eq:3l-semilin} with initial data 
\[  (s_1, s_2, h,\eta)\big|_{t=0} =
   \left\{
   \begin{array}{ll}
     (0, 1-\mu, 0, \ 1),   & \quad x<x_0 \\[2mm]
     (0, (1-\mu)/\mu, h_0, \ 0), & \quad x>x_0.
   \end{array}
  \right. \]
Note that this formulation corresponds to Cauchy problem \eqref{eq:3l-Darcy}, \eqref{eq:3l-Darcy-t0}.  Calculations on the basis of model~\eqref{eq:3l-semilin} are carried out using Nessyahu--Tadmor scheme \cite{NT90}. The results of computations are shown in Fig.~\ref{fig:fig_8} at various moments of time (solid curves). As we can see with increasing of time the solution tends to the self-similar regime (dot-dashed curves obtained by using formulae~\eqref{eq:3l-h-eta}). Moreover tip of the viscous finger propagates with the same velocity in the framework of models~\eqref{eq:3l-semilin} and \eqref{eq:3l-Darcy}.  

\section{Kinematic-wave model}
In the previous section it is shown that three-layer flow governed by the simplified 1D model~\eqref{eq:3l-Darcy} (or \eqref{eq:3l-semilin}) correctly describes the well-known fact that the fluid interface is unstable if less viscous fluid displaces more viscous one and stable otherwise.  However, the velocity of the viscous fingers propagation for these equations is the same as for the na\"{i}ve Koval model which vastly over-predicts the rate of the fingers grow \cite{Booth, Yortsos06}. A number of empirical models is proposed to reconcile this behaviour \cite{Koval, Fayers, Todd}. For example, Koval~\cite{Koval} postulates empirically that \eqref{eq:Koval} is valid if the viscosity ratio $M$ is replaced by an effective viscosity ratio $M_e$, where
\begin{equation}\label{eq:M-eff} 
  M_e= \big(M^{1/4}c_e+1-c_e)^4, \quad c_e=0.22. 
\end{equation}
Despite the fact that this model appears to work in practice \cite{Malhotra} a theoretical justification is yet to be obtained. 

As it can be seen from Fig.~\ref{fig:fig_7} and \ref{fig:fig_8}, in the framework of model \eqref{eq:3l-Darcy} or \eqref{eq:3l-semilin}, the tip of the viscous finger becomes infinitely thin with time. However, in experiments the formation of fingers of finite thickness is observed. The friction between the fluid layers is one of the possible mechanisms that prevent thinning of the viscous finger tip. Below we obtain modification of the above considered 1D models, which takes into account friction between the fluid layers. This model is new and correctly describes growth of the viscous fingers. It is proved by comparing with the Koval model (Eq.~\eqref{eq:Koval} where an effective viscosity ratio $M_e$ is used) as well as by numerical calculations on the basis of 2D Eqs.~\eqref{eq:compr-HS}, \eqref{eq:p-rho}.

\subsection{Friction between the layers}
Let us consider a two-layer flow governed by the following system of equations
\begin{equation}\label{eq:model-friction}   
  \begin{array}{l}\displaystyle 
    u_t+\beta uu_x+p_x =-\mu_1 u- \kappa\bar{\mu} (u-v)h^{-1}, \\[2mm]\displaystyle
    v_t+\beta vv_x+p_x =-\mu_2 v+ \kappa\bar{\mu} (u-v)\eta^{-1}, \\[2mm]\displaystyle
    h_t+(uh)_x=0, \quad \eta_t+(\eta v)_x=0, \\[2mm]\displaystyle
    h+\eta=1, \quad uh+v\eta=1.
  \end{array} 
\end{equation}
Here we use notations of the previous section for the layers velocities and depths; constant $\kappa>0$ is an empirical parameter, and $\bar{\mu}=\sqrt{\mu_1\mu_2}$. This system differs from Eqs.~\eqref{eq:layers} (in the case $N=2$) by the presence of additional terms with factor $\bar{\mu}\kappa$ which expresses friction on the layers interface. Note that this effect becomes significant if the thickness of one of the layers tends to zero.

\begin{figure}[t]
\begin{center}
\resizebox{0.97\textwidth}{!}{\includegraphics{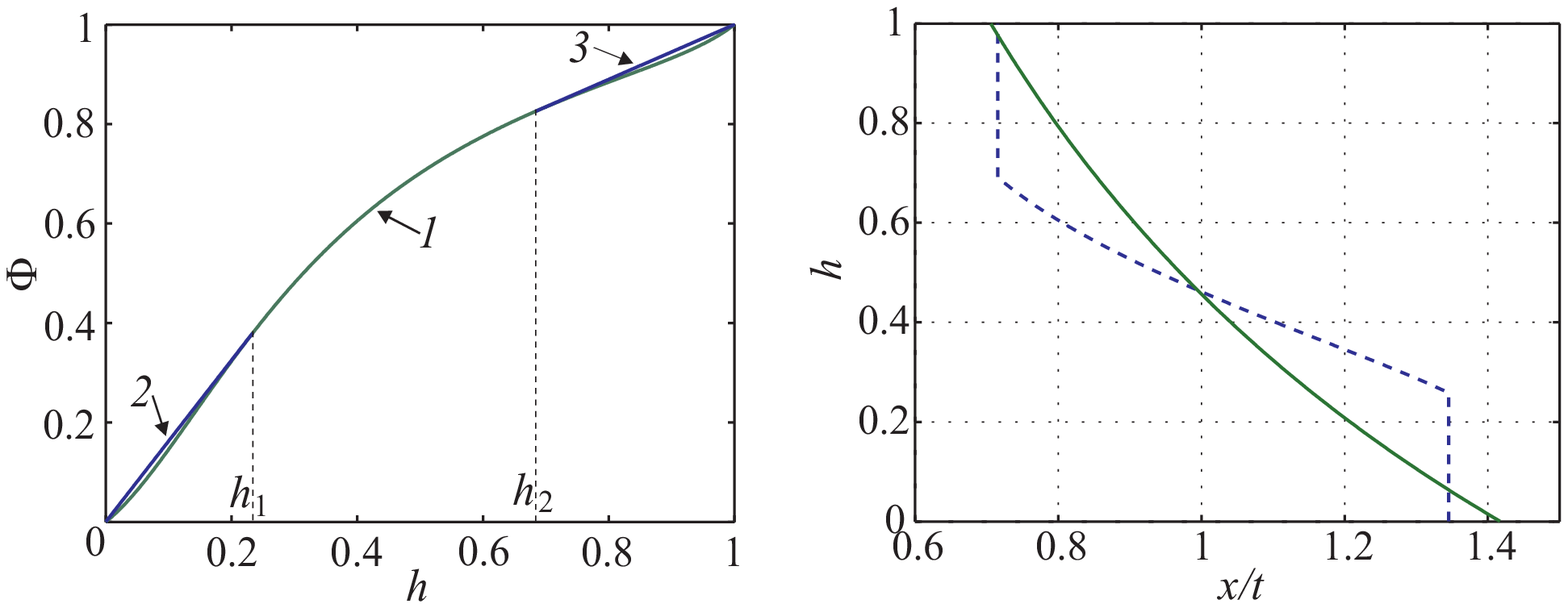}}\\[0pt]
\parbox{0.48\textwidth}{\caption{Non-convex flux $\Phi(h)$ for model \eqref{eq:kin-mod-1} with $M=10$, $\kappa=0.45$ and its convex hull.} \label{fig:fig_9}} \hfill
\parbox{0.48\textwidth}{\caption{Comparison between the Koval model (solid curve) and model \eqref{eq:kin-mod-1} with $\kappa=0.45$ (dashed) for $M=4$.} \label{fig:fig_10}}
\end{center}
\end{figure}

Further we use simplification of the momentum equations in \eqref{eq:model-friction} based on the Darcy law:
\[ p_x=-\mu_1 u-\kappa\bar{\mu} (u-v)h^{-1}, \quad 
   p_x=-\mu_2 v+\kappa\bar{\mu} (u-v)\eta^{-1}. \]
Taking into account that 
\[ \eta=1-h, \quad v=(1-uh)\eta^{-1}, \quad 
   \mu_1 u+ \kappa\bar{\mu} (u-v)h^{-1}=\mu_2 v- \kappa\bar{\mu} (u-v)\eta^{-1} \]
we obtain the following kinematic-wave model
\begin{equation}\label{eq:kin-mod-1}
 h_t+(\Phi(h))_x=0, \quad \Phi \equiv uh= \frac{(\kappa\sqrt{M}+(1-h)hM)h}{\kappa\sqrt{M}+(1-h)(1-(1-M)h)h}\,,
\end{equation}
where $M=\mu_2/\mu_1$. Typical graph of the flux $\Phi(h)$ is given in Fig.~\ref{fig:fig_9} (curve {\it 1}) for $M=10$, and $\kappa=0.45$. Obviously $\Phi(h)$ is a non-convex monotonic function such that $\Phi(0)=0$, and $\Phi(1)=1$. 

We construct self-similar solution of Eq.~\eqref{eq:kin-mod-1} with initial data $h(0,x)=1$ for $x<x_0$ and $h(0,x)=0$ for $x>x_0$. On the interval $h \in (0,1)$ function $\Phi(h)$ has two points of inflection. Thus it is necessary to construct convex hull of the flux $\Phi(h)$. For this we draw tangents to $\Phi(h)$ from the origin and from the point $(1,1)$ (lines {\it 2} ans {\it 3} in Fig.~\ref{fig:fig_9}). Let $h=h_1$ and $h=h_2$ be the tangency points (values $h_1$ and $h_2$ are obtained by solving the equations $\Phi(h_1)=h_1\Phi'(h_1)$ and $\Phi(h_2)=(h_2-1)\Phi'(h_2)+1$). According to \cite{RYa78} solution of the Cauchy problem takes the form 
\[  \xi =
   \left\{
   \begin{array}{ll}
     \Phi'(h_1), & \quad h \in [0, h_1) \\[2mm]
     \Phi'(h),   & \quad h \in [h_1, h_2] \\[2mm]
     \Phi'(h_2), & \quad h \in (h_2, 1]
   \end{array}
  \right. \]
where $\xi=(x-x_0)/t$ is the self-similar variable. This means that the solution is presented by centred rarefaction wave, which is bounded by  two ``sonic'' shocks. As we show below the numerical results of the growth of viscous fingers in the framework of Eqs.~\eqref{eq:compr-HS}, \eqref{eq:p-rho} are in good agreement with the self-similar solution of Eq.~\eqref{eq:kin-mod-1} for a suitable choice of the parameter $\kappa$.

\subsection{Comparing with the Koval model and numerical results}
Let us verify kinematic-wave model \eqref{eq:kin-mod-1} by comparing with the well-approved Koval model \eqref{eq:Koval}, where instead of $M$ effective viscosity ratio $M_e$ given by formula~\eqref{eq:M-eff} is used. We choose the following parameters: $\mu_1=2$, $\mu_2=8$. It means that the viscosity ratio $M=4$ and according to \eqref{eq:M-eff} $M_e=1.417$. Self-similar solution of the Koval model for these parameters is shown in Fig.~\ref{fig:fig_10} (solid curve). Corresponding solution of the scalar conservation law \eqref{eq:kin-mod-1} (with $\kappa=0.45$) is presented in Fig.~\ref{fig:fig_10} by dashed curve. As we can see in the framework of these models the growth rates of viscous fingers  are similar. Although the parameter $\kappa$ is a function of $M$, the values of $\kappa(M)$ are weakly vary for $M \in (1,10)$. Therefore, for the moderate viscosity ratios $M$ (less than 10) we can assume that $\kappa=0.45$. Moreover, the proposed model~\eqref{eq:kin-mod-1} describes better the structure of the finger (its thickness near the tip) than the Koval model~\eqref{eq:Koval}.

\begin{figure}[t]
\begin{center}
\resizebox{0.49\textwidth}{!}{\includegraphics{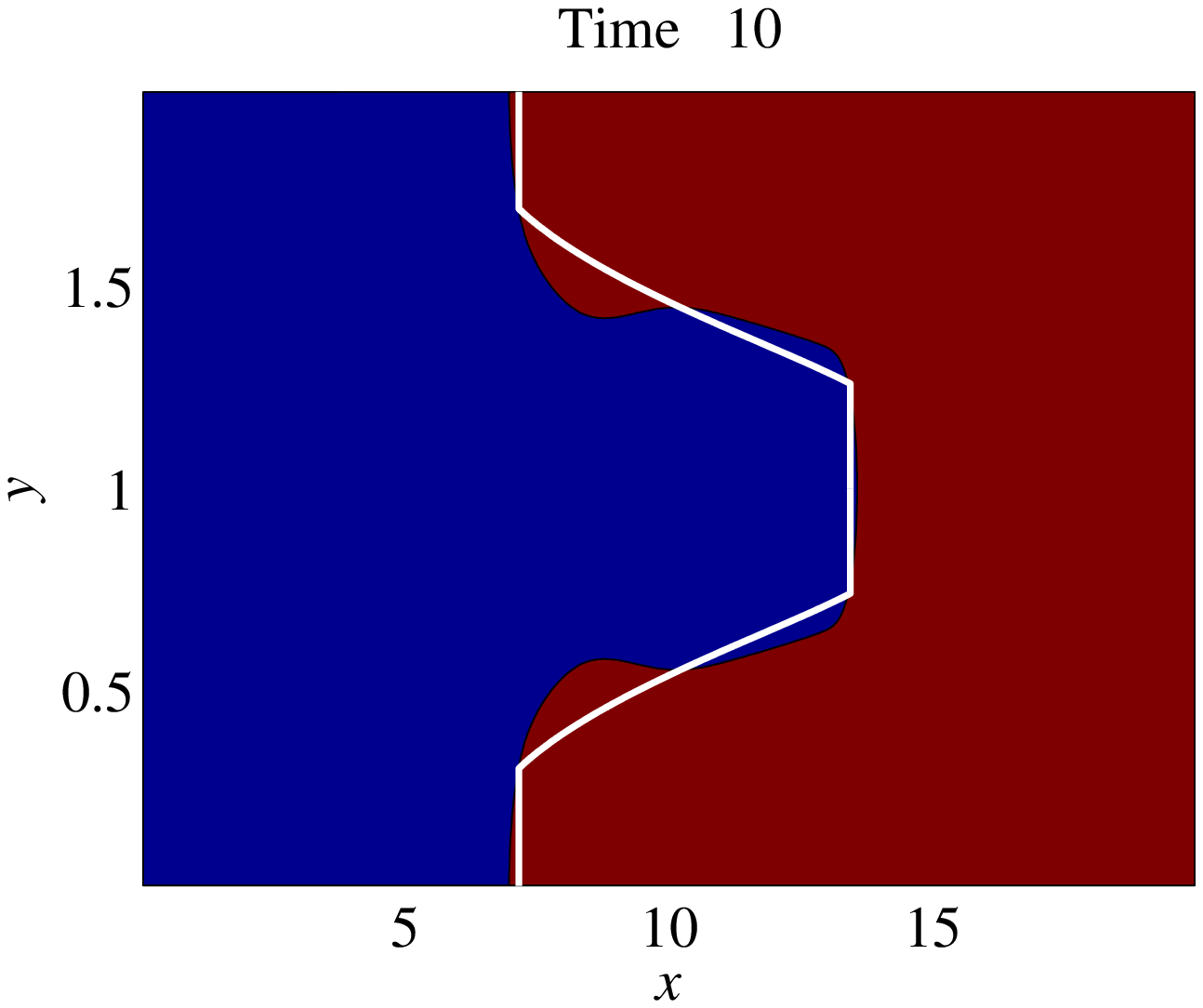}}\hfill
\resizebox{0.49\textwidth}{!}{\includegraphics{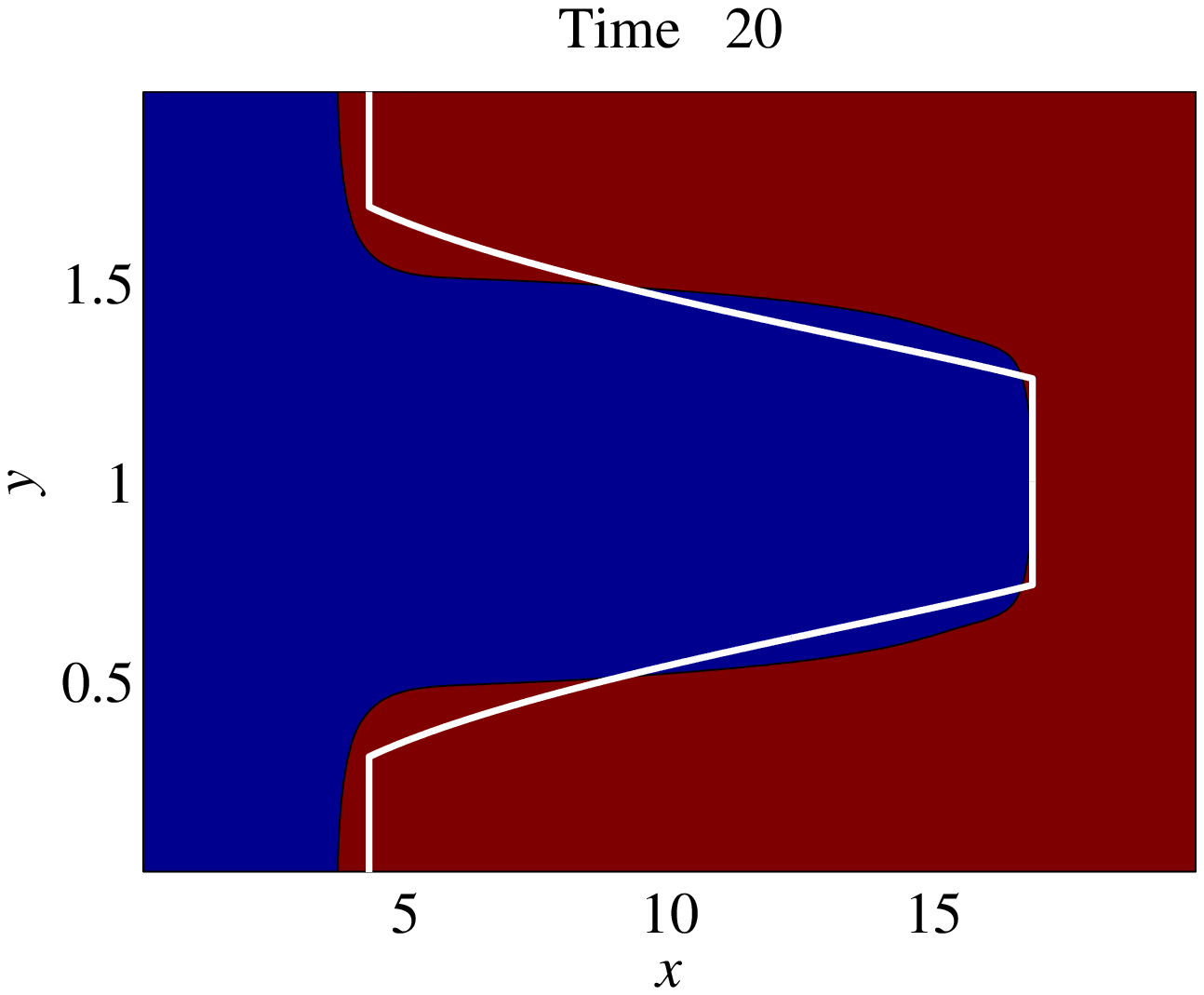}}\\[0pt]
{\caption{Comparison of the results for Eqs.~\eqref{eq:compr-HS}, \eqref{eq:p-rho} (concentration $c$ in two colour is presented) and kinematic-wave model~\eqref{eq:kin-mod-1} with $\kappa=0.45$ (white solid curve). Here we choose $\mu_1=2$, $\mu_2=8$, and $U=1$.} \label{fig:fig_11}} 
\end{center}
\end{figure}

Further we compare the results of numerical simulations of the formation of viscous fingers on the basis of 2D hyperbolic system of equations \eqref{eq:compr-HS}, \eqref{eq:p-rho} with the results obtained by using kinematic-wave model \eqref{eq:kin-mod-1}. The calculations are performed in the Hele-Shaw cell with sizes $L=20$, $H=2$ in the coordinate system moving with the average flow velocity $U=1$ with respect to $Ox$ axis. The viscosities of the fluids are equal to $\mu_1=2$ and $\mu_2=8$. At the initial time the interface $x=x_0=10$ is perturbed as follows $x=x_0+ \Delta y\, \big(\exp(-10(y-H/2)^2)-1/2 \big)$, where $\Delta y=0.2$. For discretization with respect to $x$ and $y$ we use 400 and 50 nodes respectively (calculation with finer resolution leads to the same result). On the left and right boundaries of the computational domain the reflection conditions are imposed. In 2D model~\eqref{eq:compr-HS}, \eqref{eq:p-rho} the following dependence for viscosity is used: $\mu(c)=\mu_1^{1-c}\mu_2^c$. The calculations are carried out for $\beta=1$, $\rho_0=1$ and the sound velocity $c_0=75$. In this case the change in density is not more than 0.15\%. At the same time the condition $p_y=0$ is fulfilled with high accuracy. It means that the above proposed 1D model is suitable for describing of such flow. Given above perturbation of the interface leads to the formation of a single finger which is symmetric with respect to the line $y=1$. The results of the concentration $c$ calculations using model \eqref{eq:compr-HS}, \eqref{eq:p-rho} at time $t=10$ and $t=20$ are shown in Fig.~\ref{fig:fig_11}. The distribution of $c$ is presented in two colours (blue for $c<1/2$ and brown for $c>1/2$). 

Self-similar solution of kinematic-wave model \eqref{eq:kin-mod-1} with $\kappa=0.45$ is shown in Fig.~\ref{fig:fig_11} at $t=10$ and $t=20$ by white solid lines respectively. The curves $h+1$ and $1-h$ for the model are given for a correct comparison with 2D calculations. Here self-similar variable $\xi$ is replaced by $\xi+U$. It corresponds to a transition in a moving coordinate system. The figure shows the propagation velocity and thickness of the viscous finger obtained by the kinematic-wave model agree well with the two-dimensional calculations.

\section{Conclusion} 
We derive nonlinear hyperbolic system of equations \eqref{eq:compr-HS}, \eqref{eq:p-rho} describing the flow of slightly compressible multicomponent fluid of different viscosity in a Hele-Shaw cell. On the basis of these equations simulation of jet flow and development of viscous fingers during the displacement process are performed. Calculations show that the proposed model reproduces the characteristic features of the flow associated with the development of Kelvin--Helmholtz and Saffman--Taylor instabilities (see Fig.~\ref{fig:fig_2}, \ref{fig:fig_3} and \ref{fig:fig_4}). In the case of preferential flow in the $x$-direction the pressure varies slightly in the $y$-direction that allows  to apply model of long-wave approximation \eqref{eq:model-LW} and to consider the class of layered flows described by Eqs.~\eqref{eq:layers}. Using various simplifications of the system (linearisation of the momentum equations, application of the Darcy's law) we construct a hierarchy of mathematical models of viscosity-stratified flow in a Hele-Shaw cell. These 1D models are suitable for description of the main features of the two-dimensional flow. Stationary solutions of Eqs.~\eqref{eq:layers} obtained for the three-layer flow are in good agreement with the calculations of the flow on the basis of 2D model~\eqref{eq:compr-HS}, \eqref{eq:p-rho} (see Fig.~\ref{fig:fig_6}). In the framework of the three-layer flow (systems \eqref{eq:3l-Darcy} and \eqref{eq:3l-semilin}) the interpretation of the Saffman--Taylor instability is given. Solutions of Eqs.~\eqref{eq:3l-Darcy} and \eqref{eq:3l-semilin} illustrating the formation of viscous fingers are constructed (Fig.~\ref{fig:fig_7} and \ref{fig:fig_8}). However, these models fail to predict the correct growth rate of the fingers. 

We propose modification of the layered flow model in order to agree with this behaviour. The model is obtained by including friction between the fluid layers~\eqref{eq:model-friction}. 
This provides a non-zero thickness of the fingertip and under some assumptions allows one to describe the evolution of viscous fingers on the basis of the scalar equation with non-convex flux~\eqref{eq:kin-mod-1}. Although the proposed equation~\eqref{eq:kin-mod-1} involves empirical parameter, this model reveals the physical mechanism to ensure correct propagation velocity and structure of viscous fingers. As it can be seen from Fig.~\ref{fig:fig_11} the velocity of propagation and the thickness of the fingers in proposed model~\eqref{eq:kin-mod-1} (if the  empirical parameter $\kappa$ is appropriately specified) are in fairly good agreement with calculations based on two-dimensional equations~\eqref{eq:compr-HS}, \eqref{eq:p-rho}. Comparison with the Koval model (see Fig.~\ref{fig:fig_10}) confirms the correctness of the results. 

\section*{Acknowledgements}
This work was supported by the Russian Science Foundation (grant No. 15-11-20013).

\renewcommand\baselinestretch{1}

\end{document}